\documentclass[apl,superscriptaddress]{revtex4-1}
\usepackage{graphicx}
\usepackage{natbib}
\usepackage{amsmath}
\usepackage{times}
\usepackage{epstopdf}
\begin{document}
\title{Spin torque resonant vortex core expulsion for an efficient radio-frequency detection scheme}
\author{A. S. Jenkins*}
\address{Unit\'{e} Mixte de Physique CNRS/Thales and Universit\'{e} Paris Sud, Palaiseau, France}
\author{R. Lebrun}
\address{Unit\'{e} Mixte de Physique CNRS/Thales and Universit\'{e} Paris Sud, Palaiseau, France}
\author{E. Grimaldi}
\address{Unit\'{e} Mixte de Physique CNRS/Thales and Universit\'{e} Paris Sud, Palaiseau, France}
\author{S. Tsunegi}
\address{Unit\'{e} Mixte de Physique CNRS/Thales and Universit\'{e} Paris Sud, Palaiseau, France}
\address{Institute of Advanced Industrial Science and Technology (AIST), Spintronics Research Center, Tsukuba, Japan}
\author{P. Bortolotti}
\address{Unit\'{e} Mixte de Physique CNRS/Thales and Universit\'{e} Paris Sud, Palaiseau, France}
\author{H. Kubota}
\address{Institute of Advanced Industrial Science and Technology (AIST), Spintronics Research Center, Tsukuba, Japan}
\author{K. Yakushiji}
\address{Institute of Advanced Industrial Science and Technology (AIST), Spintronics Research Center, Tsukuba, Japan}
\author{A. Fukushima}
\address{Institute of Advanced Industrial Science and Technology (AIST), Spintronics Research Center, Tsukuba, Japan}
\author{G. de Loubens}
\address{Service de Physique de l'Etat Condens\'{e} (CNRS URA 2464), CEA Saclay, Gif-sur-Yvette, France}
\author{O. Klein}
\address{Service de Physique de l'Etat Condens\'{e} (CNRS URA 2464), CEA Saclay, Gif-sur-Yvette, France}
\address{Current address: SPINTEC, UMR CEA/CNRS/UJF-Grenoble 1/Grenoble-INP, INAC, Grenoble, France}
\author{S. Yuasa}
\address{Institute of Advanced Industrial Science and Technology (AIST), Spintronics Research Center, Tsukuba, Japan}
\author{V. Cros}
\address{Unit\'{e} Mixte de Physique CNRS/Thales and Universit\'{e} Paris Sud, Palaiseau, France}
\begin{abstract}
Spin-polarised radio-frequency currents, whose frequency is equal to that of the gyrotropic mode, will cause an excitation of the core of a magnetic vortex confined in a magnetic tunnel junction. When the excitation radius of the vortex core is greater than that of the junction radius, vortex core expulsion is observed, leading to a large change in resistance, as the layer enters a predominantly uniform magnetisation state. Unlike the conventional spin-torque diode effect, this highly tunable resonant effect will generate a voltage which does not decrease as a function of rf power, and has the potential to form the basis of a new generation of tunable nanoscale radio-frequency detectors.
\end{abstract}
\maketitle 
The physics of spin-transfer torque phenomena, and most notably spin-transfer induced high frequency dynamics, has recently become a field of much experimental interest, where motion and oscillations of the magnetisation in nanostructures can be induced by the application of a spin-polarised current. These spin-torque induced excitations have been demonstrated on uniform magnetisations \cite{Kiselev2003a}, as well magnetic solitons, such as domain walls \cite{Chanthbouala2011}, magnetic vortices \cite{Pribiag2007a,Dussaux2010,Dussaux2012,Grimaldi2014,Tsunegi14} and more recently magnetic skyrmions \cite{Sampaio2013}. Interestingly, the generation of a spin-torque can be via the direct flow of a spin-polarised current perpendicular to the plane (CPP) of a magnetic tunnel junction (MTJ) or, as demonstrated more recently, by the production of pure spin currents in ferromagnet/heavy metal bilayers due to large spin orbit coupling \cite{Miron11,Liu12}.
An exciting new potential functionality for spin-torque nano-oscillators, STNOs, is in the field of high frequency detection, where a spin-torque diode induced d.c. voltage due to the mixing of the resistance and the current oscillations \cite{Tulapurkar2005} is the result of a radio-frequency (rf) current applied to the STNO. This new spintronics-based rectification effect has been shown previously for uniform magnetisation modes in magnetic tunnel junctions \cite{Tulapurkar2005,Zhu2012,Miwa2014}, with a sensitivity in excess of the conventional Schottky diode \cite{Miwa2014,FangArxiv} and is therefore of significant interest as a new type of integrated rf detector or a new generation of STNO read heads of Hard Disks. In this report, we focus on spin-torque induced resonant excitations of a single magnetic vortex confined in an MTJ. Indeed, the excitation of the fundamental mode of the magnetic vortex, i.e. the gyrotropic motion of the vortex core, has raised a lot of interest in the last decade \cite{Guslienko2002,Pribiag2007a,Dussaux2010,Grimaldi2014} where the generation of a steady state excitation of this mode has been achieved by the compensation of the damping via a d.c. current. In a more conventional approach, the vortex core can also be resonantly excited via an rf magnetic field \cite{Tsunegi14, DeLoubens2009,Uhlir2013,Vogel2010} or a rf current \cite{Jenkins2014} if the frequency of the induced core displacement resonantly matches that of the gyrotropic mode. An interesting peculiarity of the vortex-based STNOs is the ability to manipulate the chirality and polarity of the vortex \cite{Jenkins2014}, leading to the promising concept of vortex memories \cite{pigeauMem} as well as the concept of arrays of magnetic vortices used as magnonic crystals \cite{Shibata2003,Hanze2014}.

\textbf{Vortex core excitation via a spin-polarised rf current}

\begin{figure}[ht!]
\includegraphics[width=120mm]{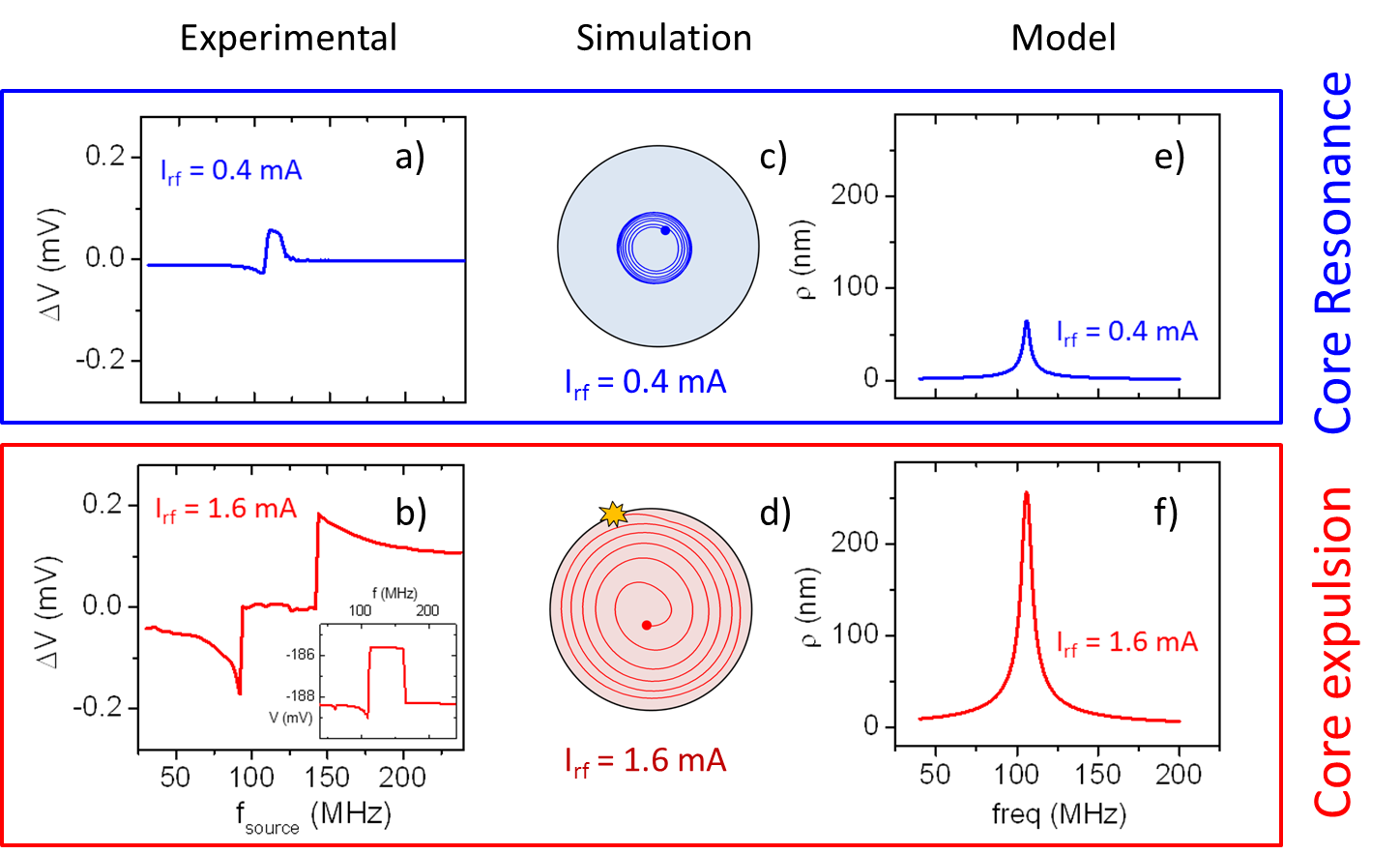}
\caption{\textbf{Vortex core resonance vs. core expulsion}. The experimentally measured voltage as a function of the source frequency for a) I$_{\text{rf}}$ = 0.4 mA and b) I$_{\text{rf}}$ = 1.6 mA, with I$_{\text{dc}}$ = 0 mA, and H$_{perp}$ = 200 mT (inset shows voltage response at I$_{\text{dc}}$ = 5 mA). The micromagnetic simulated transient trajectory of the vortex core when excited at f$_{\text{source}}$ = 100 MHz for c) I$_{\text{rf}}$ = 0.4 mA and d) I$_{\text{rf}}$ = 1.6 mA. The estimated radius of excitation, $\rho$, for e) I$_{\text{rf}}$ = 0.4 mA and f) I$_{\text{rf}}$ = 1.6 mA extracted using Eq. \ref{equation1}. }
\label{figure1}
\end{figure}

\begin{figure}[ht!]
\includegraphics[width=140mm]{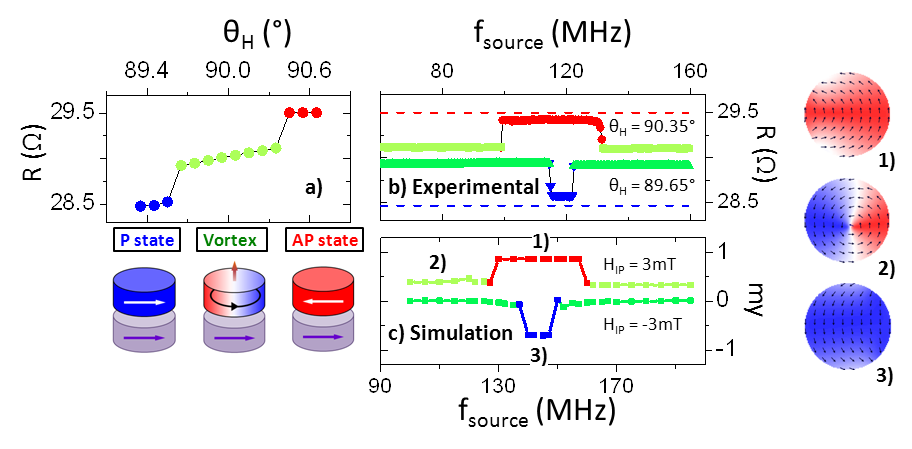}
\caption{\textbf{Core expulsion to quasi-uniform state}. a) Experimental angular dependence of the resistance showing parallel, vortex and antiparallel resistance configurations measured at I$_{\text{dc}}$ = 5 mA and H$_{\text{perp}}$ = 160 mT. b) Experimentally measured frequency dependence of the resistance for a range of rf source frequencies. The blue (red) corresponds to the parallel (antiparallel) state, and the green to the vortex state, and the vortex state is suppressed when the source is equal to the gyrotropic frequency, f$_{\text{source}}$ = 120 MHz. c) The simulated m$_{y}$ component as a function of source frequency for an in-plane magnetic field of $\pm$3 mT, where the final m$_{y}$ component is shown for the points labelled 1), 2) and 3). }
\label{figure2}
\end{figure}

One of the main purposes of this report is to present a comprehensive study of the effect of a time varying current on the magnetic vortex in order to demonstrate the exciting possibilities that exist for STNOs as a new type of nanoscale rf detector. Towards this end, we present, in Fig. \ref{figure1}, the rectified d.c. voltage which has been measured as a function of the incoming rf frequency. The magnetic tunnel junctions (diameter 500nm) we studied have a magnetic vortex as the magnetic ground state (at zero in-plane magnetic field) of the NiFe free layer with a CoFe/Ru/CoFeB synthetic antiferromagnet as the polariser layer (see Methods for details). This measurement has been performed for two rf currents, I$_{\text{rf}}$ = 0.4 mA (-22 dBm) in Fig. \ref{figure1} a), and I$_{\text{rf}}$ = 1.6 mA (-10 dBm) in Fig. \ref{figure1} b). Both these measurements have been performed with \textbf{zero} applied d.c. current and under an applied perpendicular field H$_{perp}$ = 200 mT.
For the relatively low rf current, I$_{\text{rf}}$ = 0.4 mA, a d.c. voltage, centred at f$_{\text{source}}$ = 110 MHz, is generated when the incident rf current matches that of the gyrotropic mode of the vortex. The lineshape associated with this peak in voltage can not be fit with a simple Lorentz/anti-Lorentz peaks as discussed for uniform modes \cite{Tulapurkar2005}, believed to be due to the large excitation orbit associated with magnetic vortices and the associated large nonlinearities. The generated voltage is due to the spin-diode mixing of the rf current with the TMR induced variable resistance related to the vortex gyrations \cite{Jenkins2014}. The radius of excitation associated with this peak can be estimated from Ohms law (see supplementary discussion), and was found to be $\rho_{osc}$ $\sim$ 55nm (around 20$\%$ of the total radius of the disk).
This resonant orbital trajectory of the vortex core can also be modelled micromagnetically (see Methods), and the transient trajectory of the vortex core is shown in Fig. \ref{figure1} c). Upon the application of an rf current, whose frequency is that of the gyrotropic mode, f$_{\text{source}}$ = 100 MHz, the vortex core is excited in the circular gyrotropic orbit, with a stationary radius of excitation of around 68 nm.
When the rf current is increased to I$_{\text{rf}}$ = 1.6 mA, the excitation radius of the vortex core is subsequently increased, which can lead to pronounced differences in the rectified voltage response, as shown in Fig. \ref{figure1} b). When the source frequency is close to the frequency of resonance, 100 MHz $<$ f$_{\text{source}}<$ 150 MHz, the voltage becomes zero in an abrupt transition for this measurement performed at I$_{\text{dc}}$ = 0 mA. 
This abrupt transition in voltage can be understood purely in terms of the evolution of the orbit radius of the core. When the rf current is sufficiently large, the radius of excitation approaches that of the disk, as demonstrated with the micromagnetic simulations of the transient trajectory of the vortex core, Fig. \ref{figure1} d), which show the radius of the excitation of the core will increase until it reaches the radius of the junction, at which point the core is expelled. The system subsequently enters a static quasi-uniform magnetisation state (discussed in further detail in Fig. \ref{figure2}). This vortex core expulsion explains the zero voltage state measured in Fig. \ref{figure1} b), as there is no longer a vortex present, and as such no rf current induced rectified voltage at the gyrotropic frequency. When the frequency of the incoming rf current is no longer at the resonant frequency of the vortex core, the free layer returns to the vortex state. Note that the vortex expulsion is observable in these junctions because it happens before the vortex core reaches the critical velocity for the dynamical reversal of the core \cite{Gusklienko2008}.

The variation of d.c. voltage generated by vortex oscillators can be described as $\Delta$V = $\Delta $R$_{\text{exp}}$ I$_{\text{dc}}$ + I$_{\text{rf}} \Delta $R$_{\text{osc}} \cos(\omega_{\text{S}}t)$. The first part is the d.c. voltage variation associated with the change in the mean resistance, $\Delta $R$_{\text{exp}}$, which occurs when the core is expelled. The second part is associated the conventional spin-diode effect due to the mixing of the variable resistance, $\Delta$R$_{\text{osc}}$ with the rf current, as demonstrated in Fig. \ref{figure1} a). Thus, when I$_{\text{dc}}$ = 0 mA, the voltage generated by the vortex core expulsion will be zero, as shown in Fig. \ref{figure1} b), however, when I$_{\text{dc}}$ $ \ne $ 0 mA, e.g. I$_{\text{dc}}$ = 5 mA in the inset in Fig. \ref{figure1} b), a large voltage response will be observed, which is proportional to I$_{\text{dc}}$ and highly scalable with $\Delta$R$_{\text{osc}}$ and thus with the TMR of the junction.

\textbf{Analytical modelling of the resonant core excitation}

As well as comparing the experimentally acquired results with micromagnetic simulations, we aim to further understand the observed behaviour by analytically describing the excitations of the vortex core. The force acting on the vortex core can be described as \textbf{F} = \textbf{${F_{\text{Slon}\|}}$} + \textbf{${F_{\text{Slon}\bot}}$} + \textbf{${F_{\text{FLT}\|}}$} + \textbf{${F_{\text{Oe}}}$}, where \textbf{${F_{\text{Slon}\|}}$} and \textbf{${F_{\text{Slon}\bot}}$} \cite{Slonczewski1996,Khvalkovskiy2010a,Ivanov2007,Dussaux2012} are the forces associated with Slonczewski torques from the current uniformly polarised in-plane and perpendicular to the plane, respectively, and \textbf{${F_{\text{FLT}\|}}$} is the force due to the field-like torque associated with a current polarised in-plane \cite{Zhang2002} and \textbf{${F_{\text{Oe}}}$} is the force due to the Oersted field \cite{Khvalkovskiy2009} (further details on the individual forces can be found in the supplementary materials).

For a d.c. current, the force, \textbf{${F_{\text{Slon}\bot}}$}, acts either with or contrary to the damping force felt by the core, and is responsible for the steady state oscillations mentioned previously \cite{Pribiag2007a,Dussaux2010,Dussaux2012,Grimaldi2014,Tsunegi14}. The forces due to an in-plane polarised d.c. current, \textbf{${F_{\text{Slon}\|}}$} and \textbf{${F_{\text{FLT}\|}}$}, will result in a static displacement of the core from the centre of the disk, analogous to that of an effective Zeeman field \cite{Guslienko2002}, in the $x$ and $y$ directions, for \textbf{${F_{\text{Slon}\|}}$} and \textbf{${F_{\text{FLT}\|}}$} respectively (for a d.c. current polarised along $x$).
The in-plane forces can be expressed as \textbf{${F_{\text{Slon}\|}}$} = $\Lambda_{\text{Slon}} J \textbf{u}_{x}$ and \textbf{${F_{\text{FLT}\|}}$} = $\Lambda_{\text{FLT}} J \textbf{u}_{y}$, where $\Lambda_{\text{Slon}}$ and $\Lambda_{\text{FLT}}$ are the torque efficiencies for the Slonczewski and field-like torques, respectively. At H$_{\text{perp}}$ = 0, the ratio of the efficiencies of the two forces can be expressed as $\left|\frac{\Lambda_{\text{Slon}}}{\Lambda_{\text{FLT}}}\right| = \frac{3 b}{2 R \xi_{\text{FLT}}}$, where b is the vortex core radius, R is the disk radius and $\xi_{\text{FLT}}$ is the relative efficiency of the field-like torque, and has been reported as $\xi_{\text{FLT}}$ = 0.4  for similar asymmetric magnetic tunnel junctions \cite{Chanthbouala2011}. The ratio of $b/R$ represents a factor of amplification of the field-like torque which is specific to magnetic vortices, due to the field-like torque acting over the body of the vortex, as opposed to the Slonczewski torque which will only act only upon the vortex core. The core radius is estimated from micromagnetic simulations at b $\sim$ 15 nm, i.e. around twice the exchange length, which results in the ratio $b/R$ = 0.06. As such, the field-like torque will be the dominant torque in magnetic vortices compared to the in-plane Slonczewski torque.

For a time varying current, I(t) = I$_{\text{dc}}$+I$_{\text{rf}} \cos(\omega_{S}t)$ (where I$_{\text{rf}}$ = J$_{\text{rf}}$$ \pi R^2$), the in-plane forces, \textbf{${F_{\text{Slon}\|}}$} and \textbf{${F_{\text{FLT}\|}}$}, will resonantly excite the vortex core if the incoming frequency is equivalent to the gyrotropic mode of the vortex core. The radius of these excitations can be found by introducing the rf currents into the Thiele equation (see supplementary materials) and can be expressed as:
\begin{equation}
\rho_{\text{osc}} = \frac{J_{\text{rf}} \sqrt{\Lambda_{\text{Slon}}^2+\Lambda_{\text{FLT}}^2}}{{2G \sqrt{\left( \omega-\omega_{0}\right)^2+ \frac{\left( -\pi \sigma p_{z} M_{S} L J_{\text{dc}}+D\omega \right) ^2}{G^2}}}} ,
\label{equation1}
\end{equation}
where G is the gyrovector, M$_{S}$ is the saturation magnetisation, p$_{z}$ is the perpendicular component of the spin polarisation (which is proportional to the out-of-plane magnetic field), L is the thickness of the free layer and D is the damping coefficient, J$_{\text{rf}}$ and J$_{\text{dc}}$ are the current densities for respectively the rf and the d.c. current. The total confinement is $\kappa(J_{\text{dc}}) = \kappa_{\text{ms}} + \kappa_{\text{Oe}}J_{\text{dc}}$, which is the sum of the confinement due to the magnetostatic, $\kappa_{\text{ms}}$, and Oersted, $\kappa_{Oe}$, terms, $\omega$ is the excitation frequency and $\omega_{0} = \frac{\kappa(J_{\text{dc}})}{G}  $ is the natural frequency of the vortex.

Using Eq. \ref{equation1}, we can thus estimate the orbit radius of the vortex core oscillations. For a low current, I$_{\text{rf}}$ = 0.4 mA, the radius is presented in Fig. \ref{figure1} e) as a function of the excitation frequency. The maximum excitation radius observed was 63 nm, in good agreement with that extracted experimentally ($\rho_{\text{osc}}$ $\sim$ 55 nm). For a larger rf current, our simple model predicts an orbit radius larger than the radius of the MTJ (about 260 nm for I$_{\text{rf}}$ = 1.6 mA). This new resonant behavior corresponds to the expulsion of the vortex core from the tunnel junction (see in Fig. \ref{figure1} b)) and can be well described by the analytical model.

\textbf{Final magnetization state after core expulsion }

In order to investigate the magnetic configuration of the free layer after the core expulsion, we have studied the influence of a small in-plane field, H$_{\text{IP}}$, (obtained by tilting the sample with respect to the perpendicularly applied magnetic field, H$_{\text{perp}}$). As presented in Fig. \ref{figure2} a), for H$_{\text{perp}}$ = 160 mT, when the angle at which the field is applied is $\theta_{H} <$ 89.6$^\circ$ ($\theta_{\text{H}} >$ 90.4$^\circ$), the in-plane component of the field is sufficiently strong (H$_{\text{IP}} \sim$ 3mT) that the free layer is in a uniform magnetisation state, parallel (antiparallel) relative to the polariser. For the angle range 89.6$^\circ < \theta_{\text{H}} <$ 90.4$^\circ$, the free layer is in the vortex state.

In Fig. \ref{figure2}, a d.c. current, I$_{\text{dc}}$ = 5 mA, is applied simultaneously along with the injected rf current, I$_{\text{rf}}$ = 0.8 mA, and the change in resistance when the vortex core is expelled is measured. When H$_{\text{IP}} <$ 0, i.e. $\theta_{\text{H}}$ = 89.65$^\circ$, and the source frequency is equal to the gyrotropic frequency of the core, the core is expelled to the parallel (P) uniform magnetisation state, i.e. blue dots. The opposite is true when H$_{\text{IP}} >$ 0, i.e. $\theta_{\text{H}}$ = 90.35$^\circ$, and the vortex is expelled to the antiparallel (AP) state , i.e. red dots, as shown in Fig. \ref{figure2} b). The ability to tune the final state with a small magnetic field presents the interesting concept of a three state memory, where the writing is achieved via a combination of d.c. and rf currents.

The experimentally observed behaviour is very well reproduced by micromagnetic simulations (I$_{\text{dc}}$ = 4.6 mA and I$_{\text{rf}}$ = 0.5 mA), shown in Fig. \ref{figure2} c). The simulated micromagnetic final state at three different points are presented, and show that the AP and P states (1. and 3.) are in fact C-states where the majority of the magnetisation is directed along the parallel or anti-parallel direction. This is consistent with the fact that the resistance does not quite reach the P/AP value represented by the blue/red dotted line in Fig. \ref{figure2} b).

The frequency range over which expulsion of the vortex core occurs is asymmetric for the two directions of in-plane field, which has been observed experimentally and with the micromagnetic simulations, and can be understood in terms of the shift of the vortex core by the in-plane magnetic field being offset by the shift of the vortex core due to the in-plane polarised d.c. current induced field-like torque, \textbf{${F_{\text{FLT}\|}}$} (discussed in more detail in the supplementary materials). By varying the d.c. current, the frequency bandwidth over which the expulsion is observed can effectively be tuned, and may be potentially interesting in terms of accuracy of the rf detector.

\textbf{Direct current induced compensation of damping to reduce I$_{\text{rf}}^{\text{min}}$}

\begin{figure}[ht!]
\includegraphics[width=90mm]{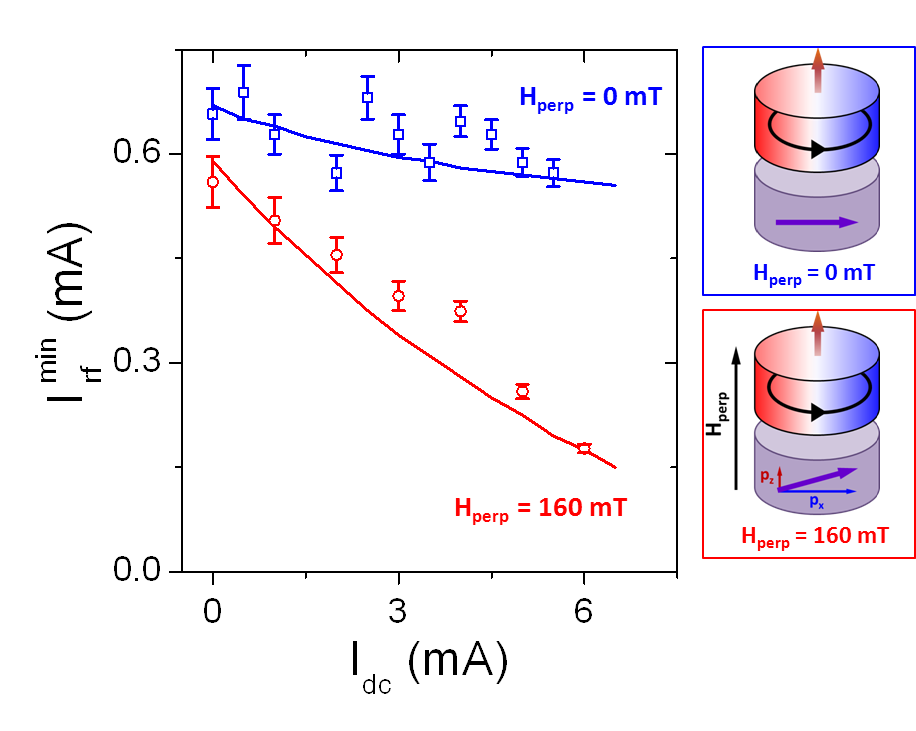}
\caption{\textbf{Reducing I$_{\text{rf}}^{\text{min}}$ by compensation of damping}. The d.c. current dependence of I$_{\text{rf}}^{\text{min}}$ for two magnetic fields, H$_{\text{perp}}$ = 0 and 160 mT and $\theta_{\text{H}}$ = 90.1$^\circ$, where the data points are the experimental data and the line is the fit with Eq. \ref{equation1}}
\label{figure3}
\end{figure}

A crucial feature for potential future applications, is the minimum rf current necessary to observe core expulsion, I$_{\text{rf}}^{\text{min}}$, which has been measured as a function of the d.c. current at two different magnetic fields, H$_{\text{perp}}$ = 0 mT (blue dots in Fig \ref{figure3}) and H$_{\text{perp}}$ = 160 mT (red dots in Fig \ref{figure3}), and the data has been fit by a model based on Eq. \ref{equation1} (see Methods).

When H$_{\text{perp}}$ = 0 mT, the magnetisation of the polarising layer is totally in-plane and as such only the effects of the in-plane component of the spin-torque terms need to be considered, i.e. \textbf{${F_{\text{Slon}\|}}$} and \textbf{${F_{\text{FLT}\|}}$}. The dependence of I$_{\text{rf}}^{\text{min}}$ is relatively constant as a function of the applied d.c. current. A small negative gradient in the d.c. current dependence of I$_{\text{rf}}^{\text{min}}$ is predicted by our model due to the combination of the d.c. current induced Oersted field altering the confinement felt by the vortex and the force due to the in-plane polarised current displacing the core from the centre of the junction (see blue line).

In the presence of an out-of-plane magnetic field, H$_{\text{perp}}$ = 160 mT, the magnetisation of the polarising layer becomes tilted out-of-plane, meaning that the out-of-plane Slonczewski term must also be considered, i.e. \textbf{${F_{\text{Slon}\bot}}$}. We emphasise that for the H$_{\text{perp}}$ which we applied, the associated spin-torque is not large enough to lead to sustained oscillations of the vortex core. The compensation of the damping due to the d.c. current induced \textbf{${F_{\text{Slon}\bot}}$} results in a larger resonant excitation radius for a given rf current, and as such leads to a strong decrease in I$_{\text{rf}}^{\text{min}}$ as a function of the d.c. current. This is particularly important for applications as it means that the power range over which a working device based on vortex core expulsion can operate can be drastically increased with the application of a perpendicularly polarised d.c. current.

\textbf{Towards applications: resonance of the core vs. core expulsion}

\begin{figure}[ht!]
\includegraphics[width=120mm]{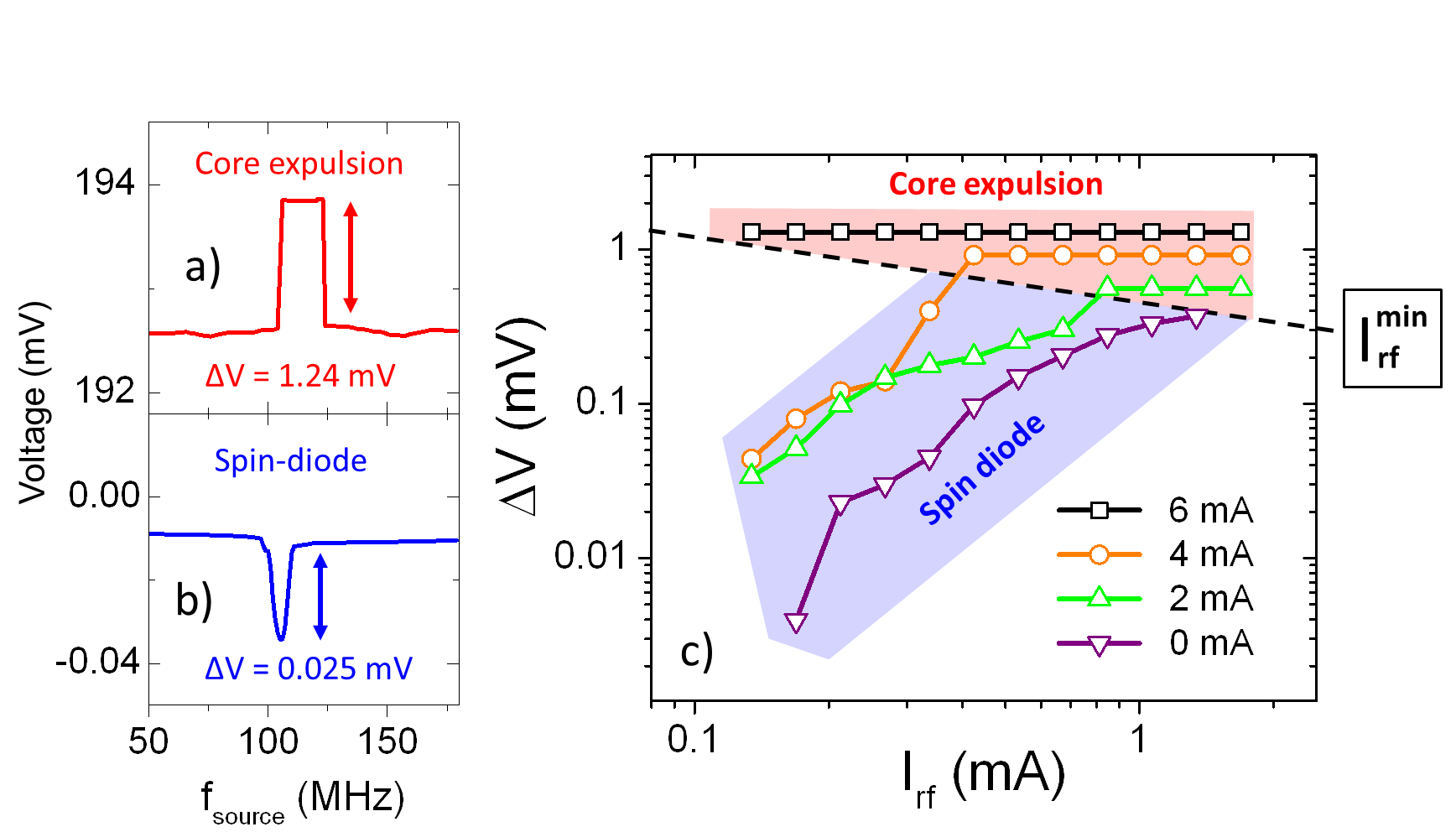}
\caption{\textbf{Radio frequency current dependence}. The rectified voltage for observed for the a) core expulsion and b) the spin-diode measurements taken at 6 mA and 0 mA, respectively, for a I$_{\text{rf}}$ = 0.2 mA. c) The generated voltage, $\Delta$V, as a function of the excitation current, I$_{\text{rf}}$, for a range of d.c. currents, at H$_{\text{perp}}$ = 150 mT and $\theta_{\text{H}}$ = 90.1$^\circ$. The core expulsion shows a constant $\Delta$V as a function of I$_{\text{rf}}$, whereas the resonant excitation decreases linearly with I$_{\text{rf}}$.}
\label{figure4}
\end{figure}

From the point of view of potential applications as a tunable nanoscale rf detector, an important feature of the core expulsion is that the when the core is expelled, the generated voltage will be constant for all values of the rf current, in contrast to the relatively smaller orbit spin-diode induced voltage, where the generated voltage is strongly reduced when the incoming rf current is reduced. To emphasise this difference, the generated voltage as a function of I$_{\text{rf}}$, for a range of d.c. currents, is presented in Fig. \ref{figure4}, for H$_{\text{perp}}$ = 160 mT. When I$_{\text{dc}}$ = 6 mA, the core is expelled for I$_{\text{rf}}$ $>$ 0.15 mA, and the generated voltage is constant for all measured values of I$_{\text{rf}}$. The value of I$_{\text{dc}}$ = 6 mA, was the maximum d.c. current that could be be applied before degradation of the tunnel barrier.
For lower values of I$_{\text{dc}}$, the vortex core is expelled if I$_{\text{rf}} >$ I$_{\text{rf}}^{\text{min}}$, and displays the characteristic constant generated voltage, but where the voltage is reduced due to the dependence on $ I_{\text{dc}}$. For smaller rf currents, I$_{\text{rf}} <$ I$_{\text{rf}}^{\text{min}}$, the spin-diode induced voltage is characterised with a sharp decrease as I$_{\text{rf}}$ decreases, as the generated voltage will depend upon the excitation radius which in turn will depend upon I$_{\text{rf}}$.
The current state of the art in terms of generated rectification voltage via the spin-torque diode effect is summarised in table \ref{table1}. The maximum (and minimum) generated voltage, V$_{\text{max}}$ (and V$_{\text{min}}$), is listed with the corresponding excitation power, P$_{\text{max}}$ (and P$_{\text{min}}$). The first two examples are taken from the literature and are for the in-plane precessional mode \cite{Tulapurkar2005} and the out-of-plane precessional \cite{Miwa2014} mode of a uniformly magnetised free layer. The last two rows, labelled Resonance (@ 0mA) and Expulsion (@ 6mA), are from this work and represent the best results observed, where the values labelled Resonance correspond to the small orbit excitations which decrease with the I$_{\text{rf}}$, as shown in Fig. \ref{figure4}.
\begin{table}[h]
\begin{tabular}{|l|l|l|l|l|}
\hline
& \textbf{V$_{\text{max}}$ ($\mu$ V)} & \textbf{V$_{\text{min}}$ ($\mu$ V)} & \textbf{P$_{\text{max}}$ (dBm)} & \textbf{P$_{\text{min}}$ (dBm)} \\ \hline
\textbf{Tulapurkar (IPP, MTJ) \cite{Tulapurkar2005}} & 32 & 3.9 & -11 & -19 \\ \hline
\textbf{Miwa (OPP, MTJ) \cite{Miwa2014}} & 1,300 & 2.4 & -26 & -67 \\ \hline
\textbf{Resonance* (@ 0mA)} & 200 & 0.63 & -5 & -32 \\ \hline
\textbf{Expulsion* (@ 6mA)} & 2,000 & 2,000 & -5 & -43 \\ \hline
\end{tabular}
\caption{Table of current state of the art in voltage rectification measurements available (* = this study)}
\label{table1}
\end{table}
The voltage which is generated from the core expulsion is independent of the excitation power (consistent with Fig. \ref{figure4}), and can clearly be seen to be greater than the previous best achieved in the literature. It is also important to note that this study was performed with MTJs which had a relatively low TMR ratio (8.5$\%$), an order of magnitude lower than the junctions presented in \cite{Miwa2014} (86$\%$). A subsequent further improvement of the maximum generated voltage associated to the vortex core expulsion may be anticipated. The maximum effective sensitivity, $\Delta$V/P$_{\text{rf}}$, possible with the vortex core expulsion in these low TMR MTJs is already as large as 40,000 V/W, compared to 12,000 V/W for the OPP modes \cite{Miwa2014}, with another order of magnitude possible with tunnel junctions with an enhanced TMR.

\textbf{Conclusion}

In summary, we demonstrate the strong interest of the excitation of a magnetic vortex confined in a magnetic tunnel junction via a resonant high frequency current both from a fundamental point of view as well as for new approaches for efficient rf detection. In particular, we show that, once the generated excitation radius of the vortex core is greater than the radius of the junction, vortex core expulsion occurs and the free layer enters the static uniform magnetisation state, leading to a large change in resistance. The rf current required in order to generate an excitation greater than the MTJ radius and therefore observe core expulsion can be reduced by displacing the vortex core laterally with an in-plane polarised d.c. current or in-plane magnetic field, or by reducing the effective damping by applying a perpendicularly polarised d.c. current. Owing to the highly scalable nature of generated voltage associated with the vortex core expulsion, magnetic vortex-based tunnel junctions make a strong case as future candidates for a new type of spintronics-based high frequency detector.

\textbf{Methods}

The data presented is for circular MTJ nanopillars with a nominal diameter of 500 nm. The stack is composed of // Synthetic antiferromagnet (SAF) / MgO (1.075) / NiFe (5) (with thickness in nm) magnetic tunnel junction. The pinned SAF layer is a PtMn (15) / CoFe (2.5) / Ru (0.85) / CoFeB (3) multilayer. The NiFe layer is the free layer with a magnetic vortex as the ground state. Standard d.c. and high frequency current sources are used to source I$_{\text{dc}}$ and I$_{\text{rf}}$, respectively, and the voltage measured with a digital voltmeter connected via a bias tee. The perpendicular component of the spin polarisation is estimated at p$_{z}$ = 0.15 at H$_{\text{perp}}$ = 200 mT, from the tunnelling magnetoresistance field dependence. A positive d.c. current polarised in the p$_{z}$ direction will act to compensate the damping felt by the vortex core. Before each measurement was made, a large perpendicular magnetic field and dc current were applied to reliably initialise the vortex core polarity and chirality, before the frequency of the rf source was swept around the gyrotropic frequency of the vortex. The measurements were performed on multiple samples (all of similar composition and size) and the same results were obtained Sample 1 (R$_{P}$ = 28.5 $\Omega$) was used for the angle dependence presented in Fig. \ref{figure2}, and Sample 2 (R$_{P}$ = 30.9 $\Omega$) was used for the d.c. current and I$_{\text{rf}}$ dependence presented in Fig. \ref{figure4}, as well as Fig. \ref{figure1}.
Micromagnetic simulations were carried out with the OOMMF code \cite{OOMMF}. Standard values for NiFe were taken, M$_{S}$ = 580x10$^3$ A/m, Gilbert damping $\alpha$ = 0.009, and the exchange stiffness A = 1.3x10$^{-11}$ J/m. The mesh size was 5x5 nm, for a thickness of 5 nm. The polarisation was P = 0.2 and a field-like torque efficiency $\xi_{\text{FLT}}$ = 0.4. The vortex core was initially located in the centre of the 500 nm diameter disk and was excited with a current polarised in the (0 0.8 0.2) direction.
The radius was modelled by calculating $\rho$ = $\rho_{\text{osc}}$ + $\rho_{\text{disp}}$ + $\rho_{\text{H}}$, and expulsion is assumed when $\rho > $0.75R, as observed in micromagnetic simulations. The parameters used were M$_{S}$ = 580x10$^3$ A/m, $\alpha$ = 0.009, and A = 1.3x10$^{-11}$ J/m for a polarisation of 0.2 and a $\xi_{\text{FLT}}$ = 0.4, which allowed a close correlation between the model and the experiment in terms of generated frequency, and d.c. and rf current dependence.

\textbf{Acknowledgements}
The authors acknowledge the ANR agency (SPINNOVA ANR-11-NANO-0016) as well as EU FP7 grant (MOSAIC No. ICT-FP7-8.317950) for financial support. E.G. acknowledges CNES and DGA for their support.

\textbf{Author Contributions}
A.S.J., P.B. and V.C. conceived and coordinated the project. A.S.J. performed the experimental measurements, as well as the micromagnetic simulations. A.S.J, R.L, E.G., P.B. and V.C. interpreted the data. S.T. assisted in the development of the experimental set-up. The samples were fabricated by H.K., K.Y., A.F., S.Y. The manuscript was prepared by A.S.J with the assistance of P.B., V.C. All authors commented the manuscript.

\bibliographystyle{unsrt}

\end{document}